# ON THE LINK BETWEEN OPTOACOUSTIC IMAGING AND SELECTIVE PHOTOTHERMOLYSIS

Authors: Sergio Contador[1,2], Rodrigo Rojo[1,2], Alvaro Jimenez[1,2], Juan Aguirre[1,2]*

Affiliations:

1. Department of Electronic and Communications Technology, Medical Engineering Development and Innovation Center, Autonomous University of Madrid, Madrid, Spain.

2. Health Research Institute of the Jiménez Díaz Foundation, Madrid, Spain.

*Corresponding author: juan.aguirre@uam.es

## Abstract

Selective photothermolysis (SP) is widely used in clinical and cosmetic dermatology to remove unwanted skin structures. Careful laser parameter selection results in safe and effective target removal. Nevertheless, parameter selection relies on a trial-and-error process based on visual inspection of the immediate skin response. This process is highly dependent on the practitioner experience and can be time-consuming.

SP and optoacoustic imaging (OI) share many physical principles. However, the possibility of using OI to improve laser parameter selection in SP has not been studied before. Here, we explore the relationship between OI and SP theoretically and through clinical in-human trials with a focus on tattoo removal.

Our results demonstrate a strong correlation between OI signals acquired before and after treatment with the immediate clinical endpoint, suggesting that OI could be used as a tool for optimal parameter selection and reduced treatment duration in tattoo removal and other SP treatments.

# Introduction

In 1983 Anderson and Parrish proposed selective photothermolysis as a tool to selective damage structures, cells, or organelles that exhibit a high light absorption coefficient compared to its surrounding tissue[1]. To perform SP, the tissue is illuminated using a very short pulsed laser beam at a wavelength that ensures absorption by the target and not by the surrounding tissue. If the energy is high enough the target is damaged through heat deposition. The short pulse length ensures minimal collateral effects on surrounding tissue due to the so-called "heat confinement effect". More in detail, the short laser pulse ensures that the deposited heat remains confined to the tissue regions that absorbed the light, since heat will diffuse effective only during the time the excitation light is "on".

For the specific tattoo removal case, tissue is illuminated using a pulsed laser beam at a specific wavelength and energy. Sufficient energy per pulse (~500 mJ) causes fragmentation of the ink granules, which are subsequently removed by the immune system. Selecting the appropriate wavelength ensures that the target molecules absorb the laser energy, rather than the surrounding tissue. This heat deposition comes with a temperature increase and mechanical shock, which ultimately fragments the pigments into smaller granules. The laser-induced shock allows the immune cells to clear away the ink. The procedure is repeated multiple rounds, spaced in several weeks, to ensure complete removal of the pigments[2].

Selection of the appropriate laser parameters i.e per pulse energy and wavelength is crucial[3] (see discussion for information regarding spot size). To determine them dermatologists, perform a trial-and-error procedure. Firstly, wavelength is selected as a function of the colour of the tattoo (e.g 1064nm for black, 730 nm or 785 nm for blue and green). Then, the practitioner observes the laser first-pass reaction of a small area of the tattoo[4]. The objective is to obtain a temporary whitening due to dermal gas bubble formation with the lowest energy per pulse possible. The whitening is used as an immediate treatment endpoint since it indicates that the tattoo has absorbed light and got heated up violently. Importantly, a too strong whitening effect will result in excessive damage to the skin. However, a non-expert practitioner might not be able to adjust correctly the whitening level. Moreover, even expert practitioners might not be able to distinguish whether the whitening of the skin is dermal (i.e. due to tattoo heat) or epidermal (i.e. due to the destruction of superficial melanocytes)[4], resulting in a high risk of unwanted skin lesion and no removal of the tattoo. Lastly, this method might be too lengthy for tattoos with multiple colours, especially if the laser systems have 4 or 5 wavelengths available, resulting in suboptimal wavelength or energy selection.

Practitioner may also apply the so-called "R20" method [5]. In this procedure, once the optimal laser parameters are obtained, the operator applies multiple lasers passes during the same session instead of waiting for weeks to perform a new pass. To this end, the dermatologist must wait for ~20 minutes after the first pass, to ensure that whitening is gone, since the "frosting" prevents the laser light to reach the tattoo due to increased photon scattering. However, 20 minutes is very a long time, resulting in inefficient management of clinical costs.

As an absorption-based interrogation method, OI exploits the conversion of light transients in thermal energy[6–9] sharing many of the mechanism of SP. More specifically, in optoacoustics, tissue is illuminated with laser light with a pulse duration tuned for achieving the heat confinement condition, like in tattoo removal. However, the energy per pulse is only few tens of microjoules and the stress confinement condition must be also met. Upon light absorption by chromophores, heat is deposited leading to a slight temperature increase, inducing thermoelastic expansion, which creates stress and finally an optoacoustic wave. The optoacoustic wavefront is then detected with transducers placed outside tissue. After signal digitization an image representing the light absorption map can be obtained through mathematical processing of the OI data. Importantly, the depth to resolution ratio is very high since it is determined by ultrasound diffraction, enabling imaging major dermal structures [10–15]

Because of these unique features, optoacoustic interrogation may provide not only morphological images showing the depth and width of the tattoo due to its favourable penetration to depth ratio, but also dosimetric images, since its contrast mechanism is light absorption/heat deposition. However, no studies have yet explored how optoacoustic imaging could address the challenges associated with tattoo removal described above.

In this study we explore the theoretical relation between OI and SP. Moreover, we have performed, for the first time, clinical studies to establish the correlation between the clinical endpoint in tattoo removal and the OI signals obtained before and after a laser pass. To this end we used the recently developed SOL system [16]. Finally, in light of the results we discuss how OI can be used to improve SP tattoo removal in terms of optimal wavelength and energy selection, reduced laser parameter selection time and reduced of the duration of the R20 method.

# Methods

## Theoretical analysis of the relations between optoacoustic imaging and selective photothermolysis

### *Optoacoustic imaging*

To perform optoacoustic imaging, tissue is illuminated with low-energy pulsed light, typically ranging from 10 μJ to a few mJ per pulse, depending on the application. As light diffuses through in tissue (Figure 1a) it is preferentially absorbed by structures with higher optical absorption coefficients, such as ink granules (Figure 1b), leading to heat deposition. If the heat confinement condition is met, the following expression holds[17]:

$$H_{OA}(r) = \Delta T(r)\rho c_v \qquad (1)$$

Where $H_{OA}(r)$ is the deposited heat per unit volume, $\rho$ is the mass density, $c_v$ is the specific heat and $\Delta T(r)$ is the temperature rise typically on the order of 0.01 ºC. Equation 1 implies that the light energy is transferred to the tissue locally, resulting in a local increase in temperature without significant heat diffusion.

The increment of temperature lead to tissue stress. If the stress-confinement condition is met (that is, the stress does not have time to travel away from the absorbing structure), the following expression holds:

$$\Gamma\, H_{OA}(r) = p_0(r) \qquad (2)$$

where $\Gamma$ is the dimensionless Grünneisen parameter, and $p_o(r)$ is the initial optoacoustic pressure. Equation 2 implies that the transfer of heat into pressure is maximized being weighted by $\Gamma$ . The initial pressure then propagates through tissue as an optoacoustic wave (Figure 1c). After detection and digitization of the pressure signals, an image representing $p_o(r)$ can be obtained[17] using image formation algorithms (Figure 1h). It is important to note that the stress confinement condition is much more restrictive that the heat confinement condition, since stress travels much faster that heat. For example, for a 10-µm-diameter ink sphere, the stress-confinement condition is satisfied for laser pulse durations shorter than approximately 6–7 ns, whereas the heat-confinement condition is satisfied for much longer durations, typically on the order of tens of microseconds, depending on the thermal diffusivity assumed.

***Selective photothermolysis***

To perform selective photothermolysis, tissue is illuminated with short picosecond to nanosecond pulses (Figure 1e), as in optoacoustic imaging. However, the energy per pulse is orders of magnitude larger (500 mJ to 1J). As light travels through tissue (Figure 1a) it is absorbed preferentially by the structures with higher optical absorption coefficients (e.g. the ink granules, Figure 1f), depositing heat. If the heat confinement condition is met the following expression holds[1]:

$$H_{sp}(r) = \Delta T(r)\rho c_v \tag{3}$$

As in the optoacoustic case, $H_{SP}(r)$ is the deposited heat per unit volume, $\rho$ is the mass density, $c_v$ is the specific heat, and $\Delta T(r)$ is the temperature raise. Again, equation 3 indicates that heat is transferred into the tissue "locally" resulting in a local increment of temperature with no heat diffusion. However, unlike in OI the temperature rise is so large that the ink particles are fragmented through thermal effects, mechanical stress and intense optoacoustic waves. In addition, temporary frosting occurs due to the formation of subepidermal gas bubbles (Figure 1g-h).

Stress confinement is not a sine-qua-non condition for SP. Nevertheless, satisfying the stress-confinement condition would result in a stronger optoacoustic effect. As a result, the generated optoacoustic wave would play a more dominant role in ink fragmentation.

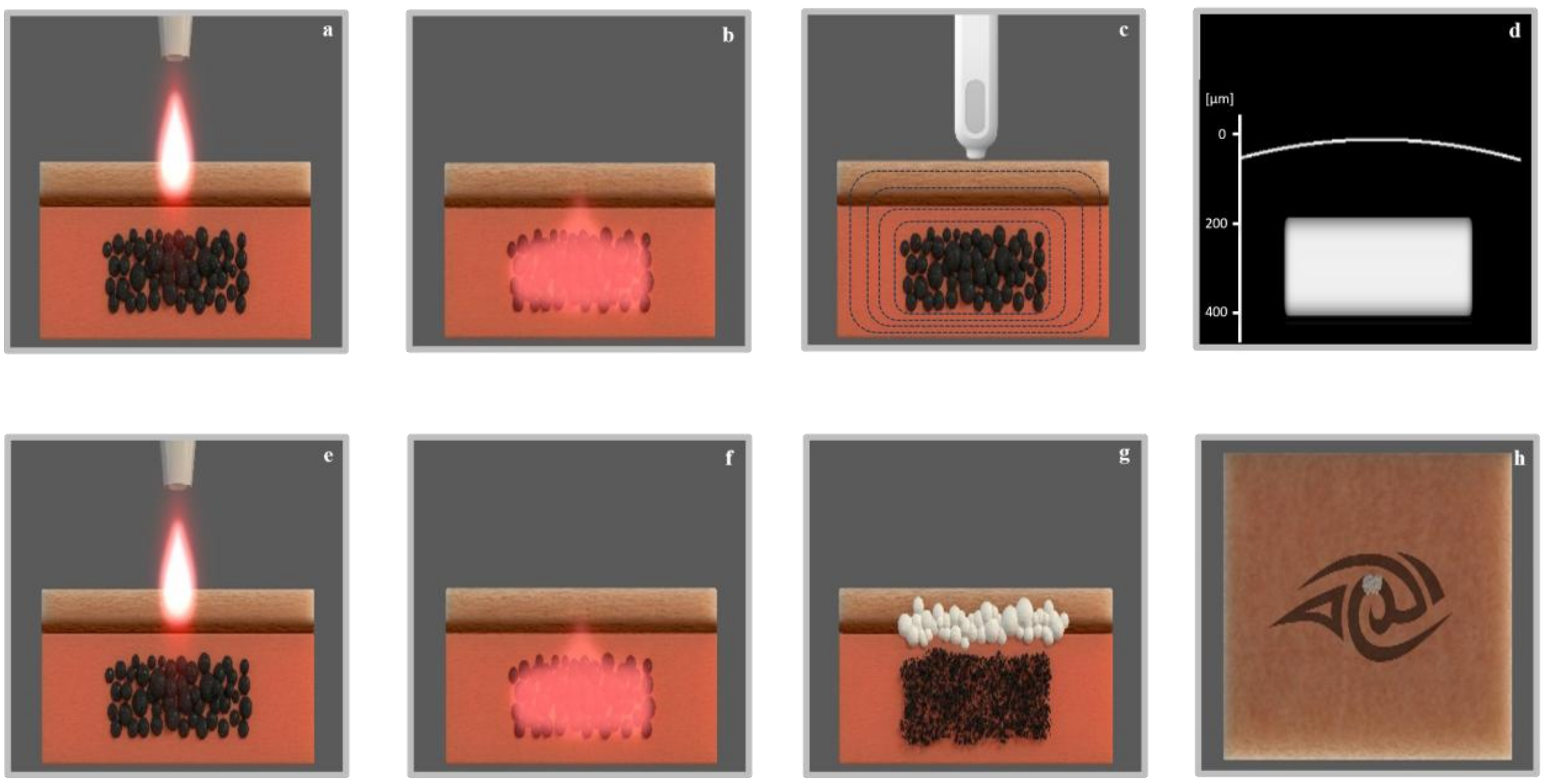

**Figure 1**: Conceptual relationship between SOL optoacoustic imaging system and selective photothermolysis for tattoo interrogation. (**a-d**) Description of the SOL imaging system. (**a**) In SOL imaging, the skin is illuminated with a short, low-energy light pulse (0.1-10 ns; ≈0.01 mJ). (**b**) The light is preferentially absorbed by the tattoo pigment. (**c**) Light absorption induces thermoelastic expansion and generates an optoacoustic wave that is detected by the ultrasound transducer. (**d**) A representative depth-resolved SOL image reconstructed from the detected optoacoustic waveform, showing the optical absorption distribution of the tattoo layer. (**e-h**) Description of selective photothermolysis. (**e**) Tattooed skin is illuminated with a short, high-energy light pulse (0.1-10 ns; ≈100 mJ). (**f**) When the wavelength and spot size are appropriately selected, light is preferentially absorbed by the tattoo pigment. (**g**) Because of the short pulse duration, the deposited heat remains confined to the pigment zone and leads to the characteristic whitening (clinical endpoint) observed after laser exposure. (**h**) Representative surface appearance of the tattooed skin.

### *Optoacoustics can complement SP*

In an idealized physical scenario, the temperature increase of ink granules during an SP process could be estimated as a function of SP laser energy from the intensity of non-invasive optoacoustic images. This prediction could be achieved by combining the expression $H = \mu\vartheta$ with Equations 3 and 2, where μ is the ink absorption coefficient and θ is the fluence. Such an estimation would assume that the illumination pattern is identical for both SP and optoacoustic imaging, that the laser beams differ only in energy, and that Γ is known. By predicting the temperature increase, the treatment energy or wavelength could be adjusted with high precision.

However, a more realistic physical model indicates that Equation 3 only holds in the so-called linear regime. If the temperature increase is too high, other heat-transport phenomena come into play through phase changes and other physicochemical reactions. The mechanical processes involved in fragmentation also undermine the validity of Equation 3. Nevertheless, there is still a direct, yet complex relation between the intensity of the optoacoustic images/signals and the amount of deposited heat in a subsequent SP process.

In any case, equation 1,2 and 3 indicates that optoacoustic images provide information on where the laser will be preferentially absorbed when performing an SP treatment, provided that the wavelengths of the optoacoustic system and the SP systems are the same.

Importantly, a multispectral optoacoustic imaging system would provide information about the wavelength that most efficiently deposits heat in the ink, allowing optimization of wavelength selection in SP.

### The SOL optoacoustic imaging system

In order to contrast our theoretical analysis experimentally, we have used the SOL imaging system (Figure 2). SOL is a compact handheld platform designed for optoacoustic imaging of layered biological structures. A detailed technical description has been reported elsewhere by Contador et.al.[16]; here, only the main features are summarized.

The system consists of four main elements: an optical excitation subsystem, a custom-made handheld probe, a data acquisition unit (DAQ), and a laptop for signal processing and visualization (Figure 2a). Optical excitation is provided by a Cobolt Tor XS Q-switched laser (HÜBNER Photonics, Solna, Sweden) at 1064 nm, delivering short pulses in the microjoule range through a multimode optical fiber FP1000ERT (Thorlabs) with 1 mm core diameter integrated into the probe. The probe also incorporates a single element flat piezoelectric transducer V324-SM (Olympus NDT; Waltham, MA, USA), with a central frequency of 25 MHz, enabling co-localized optical delivery and acoustic detection directly on the skin surface (Figure 2b). The acquired signals are digitized by a PicoScope 3406D oscilloscope (Pico Technology) at a sampling rate of 500 MS/s over a 7 µs window, and transferred to a Hewlett Packard ProBook 640 G8 laptop (HP Inc., Palo Alto, CA, USA), equipped with a 4.2 GHz processor and 8 GB of RAM, for real-time processing and visualization using custom PYTHON routines.

In contrast to tomographic optoacoustic systems, which require multiple detection positions and computationally demanding image reconstruction[18–20], SOL uses a simplified non-tomographic image formation strategy tailored to the interrogation of layered absorbers such as tattoo ink[16]. Briefly, each recorded optoacoustic waveform is converted from time to depth using the speed of sound in tissue. Thanks to the coplanar geometry of the transducer and the imaged target (tattoo ink), the signals incoming from the layered object are highly amplified while the signals incoming from other objects are dumped. The relevant portion of the signal can be then isolated by a simple signal thresholding. Next, the signal envelope is obtained using the Hilbert transform and corrected for baseline drift before visualization. The resulting one-dimensional depth-resolved profile is laterally expanded into a grayscale strip image, smoothed, and normalized for display as a representative depth-resolved optical absorption image (Figure 1h). This image formation approach is well suited for localizing dominant layered absorbers and monitoring relative changes in optoacoustic signal amplitude, which is the main parameter analysed in this study.

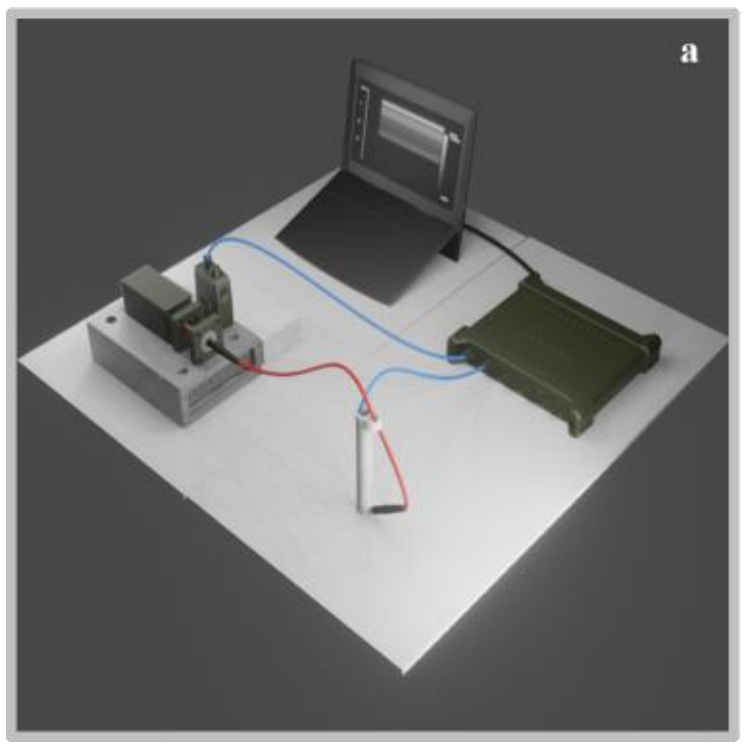

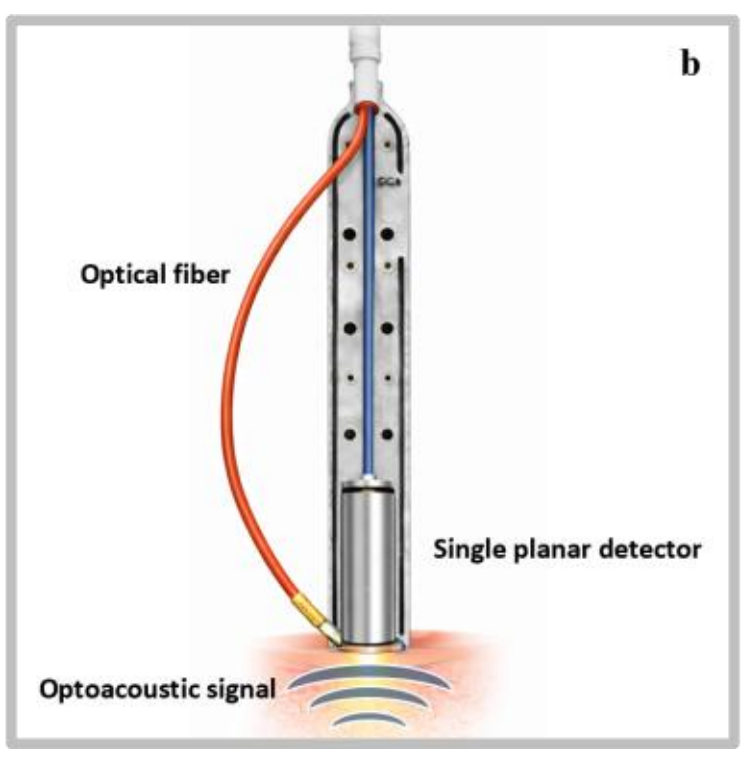

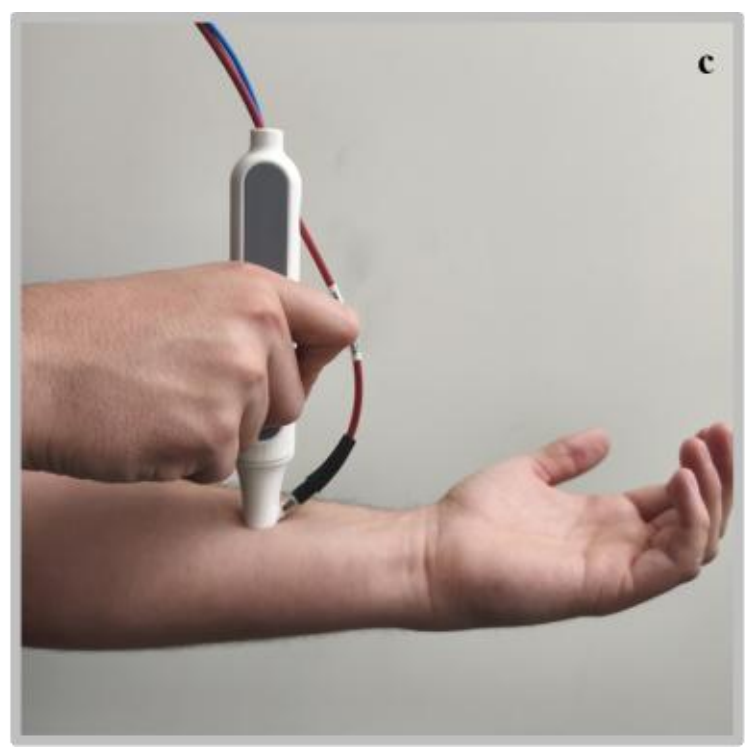

**Figure 2**: The single-layer optoacoustic (SOL) imaging system. (**a**) Schematic representation of the custom-built SOL system for tattoo interrogation, consisting of four main components: the optical excitation subsystem (left), the handheld probe (bottom), the DAQ unit (right), and a laptop (top) used for signal processing and visualization. (**b**) Schematic of the handheld probe, showing the integrated planar ultrasound transducer and the optical fiber used for light delivery, together with the optoacoustic signal coming from the tattoo to the detector. (**c**) Photographic view of the handheld probe used for in-vivo measurements.

**Clinical study 1: Relation between the clinical endpoint and OI signal obtained before a laser pass**

A clinical pilot study was conducted to evaluate whether the optoacoustic response measured with the SOL system, before treatment, is correlated with the immediate clinical endpoint of selective photothermolysis, as predicted by the theoretical analysis. A positive correlation can lead to possible improvements in the energy and wavelength selection in SP (see discussion).

Ten tattooed skin zones and two non-tattooed skin zones were included in the study. The analysed tattooed zones were located on the right leg, both arms, and the left back, and included one yellow zone, one red zone, and eight black ink zones. For each zone, optoacoustic data were acquired with the SOL system before laser treatment. For each zone, laser treatment was delivered using a 1064 nm picosecond Nd:YAG laser system (Pico Clear; Alma Lasers Ltd., Caesarea, Israel) with a fluence of 1.1 J/cm², a repetition rate of 3 Hz, and a spot diameter of 6 mm. One photographic picture was obtained before treatment, and a second photographic picture was obtained immediately after treatment to document the skin response.

Before treatment, five optoacoustic measurements were acquired for each zone. The optoacoustic response was quantified using the peak-to-peak amplitude (P2P) of the recorded signals. For each measurement, a mean P2P value was calculated from the five captures and used for subsequent analysis.

Immediately after treatment, each zone was classified into two categories according to the observed whitening response: no whitening or whitening. Normality of the P2P values was assessed using the Shapiro–Wilk test. Since normality was not met and only two independent categories were compared, differences between the no whitening ($n = 20$ P2P values) and whitening ($n = 40$ P2P values) groups were evaluated using the Mann–Whitney U test. Statistical significance was defined as $p < 0.05$, and values were reported as mean ± standard deviation. All reported p values were two-sided.

**Clinical study 2. Relation between the clinical endpoint and OI signal obtained after a laser pass**

A second clinical pilot study was conducted to investigate whether the SOL signal obtained after treatment can be related with the evolution of the clinical response. We expect the optoacoustic signal to recover as the whitening effect disappears. The disappearance of the whitening imply that the scattering reduces and the excitation light as long as the ink remains in place. Eventually, scattering is low enough for the light to reach the tattoo. A clear correlation can lead to possible improvements in the length of R20 measurements (see discussion).

A total of four black ink zones were included, located on the right leg, right arm, left ankle, and left side of the back. For each zone, baseline optoacoustic data and one photographic picture were acquired immediately before laser exposure. Treatment was then delivered using the same platform used in "Clinical study 1", with the same repetition rate and spot diameter. Laser fluence varied across participants according to the criteria of an expert in tattoo removal: 4.1 J/cm² in two zones, and 2.8 J/cm² or 5.1 J/cm² in one zone each. Following treatment, one photographic picture and one optoacoustic measurement were acquired every 3 min from 3 to 30 min after treatment from each zone. As in "Clinical study 1", each optoacoustic measurement consisted of five captures, and a single mean P2P value was assigned to each measurement and used for subsequent analysis. Thus, 11 measurements and 55 individual captures were collected per zone. For longitudinal visualization, P2P values were normalized to the corresponding baseline value obtained before treatment in each zone.

To assess the temporal evolution of the optoacoustic signal in the context of the R20 method, measurements obtained at 3, 6, 9, 12, 15, and 18 min after treatment were compared with the reference measurement acquired closest to 20 min after treatment (i.e. 21 min).

The tattoo ink might disappear right after the laser pass, in this case we don't expect any recovery of the SOL signal. To assess this effect, we conducted a statistical test. Measurements acquired from 6 to 30 min after treatment (n = 36 P2P values) were analysed as a function to the percentage of visible tattoo pigment removal within the treated region. The relationship between tattoo pigment removal percentage and optoacoustic response was analysed by fitting a linear regression model. Normality was assessed using the Shapiro–Wilk test. Since normality was met, an ANOVA F-test of the regression model was used to determine whether tattoo pigment removal significantly explained the variability in the optoacoustic response. Statistical significance was defined as $p < 0.05$.

In the linear regression model, the dependent variable was the mean P2P amplitude measured in each treated zone at each time point from 6 to 30 min after treatment, whereas the independent variable was the percentage of tattoo pigment removal calculated for the corresponding treated zone.

The percentage of tattoo pigment removal was calculated from the photographic pictures acquired at 3 and 30 min after treatment. The treated region of interest (ROI) was defined from the photographic picture acquired 3 min after treatment, when the laser-induced local reaction was clearly visible due to the immediate whitening response. The ROI contour was manually delineated in red and subsequently converted into a filled internal ROI mask. Tattoo pigment removal was then quantified from the photographic picture acquired 30 min after treatment. The image was converted from RGB to the CIELAB colour space, and the L* channel was used to classify pixels according to lightness differences between tattoo pigment and tattoo pigment removal. This channel was selected because, in the CIELAB colour space, L* encodes luminance/lightness separately from the chromatic components a* and b*. Therefore, it provided a direct intensity-based criterion for distinguishing darker pigment from lighter tissue areas corresponding to tattoo pigment removal. Within each ROI, two 5 × 5 pixel reference regions were selected, one representative of tattoo pigment and one representative of tattoo pigment removal. The median L* value was calculated for each reference region, and the classification threshold was defined as the midpoint between both median values. ROI pixels were then classified as tattoo pigment or tattoo pigment removal according to this threshold. Finally, the tattoo pigment removal percentage was calculated as the proportion of analysed ROI pixels classified as tattoo pigment removal relative to the total analysed ROI.

# Results

**Clinical study 1.**

Representative examples of the immediate whitening response are shown in Figure 3a-c, corresponding to whitening (black ink) and no whitening (yellow ink, red ink, and skin) categories. The whitening effect was clearly visible in the black ink zones, whereas in the yellow/red ink zones or in skin, it was not visibly observed. The mean Hilbert envelope optoacoustic profiles shown in Figure 3d were computed using all baseline measurements available within two whitening categories and are displayed together with their standard error of the mean (SEM) bands. These profiles revealed clear differences between both categories. Zones classified as whitening exhibited a marked increase in signal amplitude, with a prominent peak at the tattoo depth, whereas zones with no visible whitening showed low amplitude responses.

Quantitative analysis, summarized in the boxplot in Figure 3e, confirmed this trend. The two boxplots are clearly separated and show no overlap. The mean P2P value was 5.01 ± 1.18 a.u. for the no whitening category and 43.72 ± 26.91 a.u. for the whitening category. Thus, baseline optoacoustic amplitude increased with the appearance of the immediate whitening response. A significant difference in P2P was found between the no whitening and whitening categories, with the whitening category showing significantly higher P2P values than the no whitening category (Mann-Whitney U = 0, $p < 1 \times 10^{-6}$).

Overall, these results indicate that higher optoacoustic responses before treatment were associated with a stronger immediate clinical endpoint (whitening) after laser exposure.

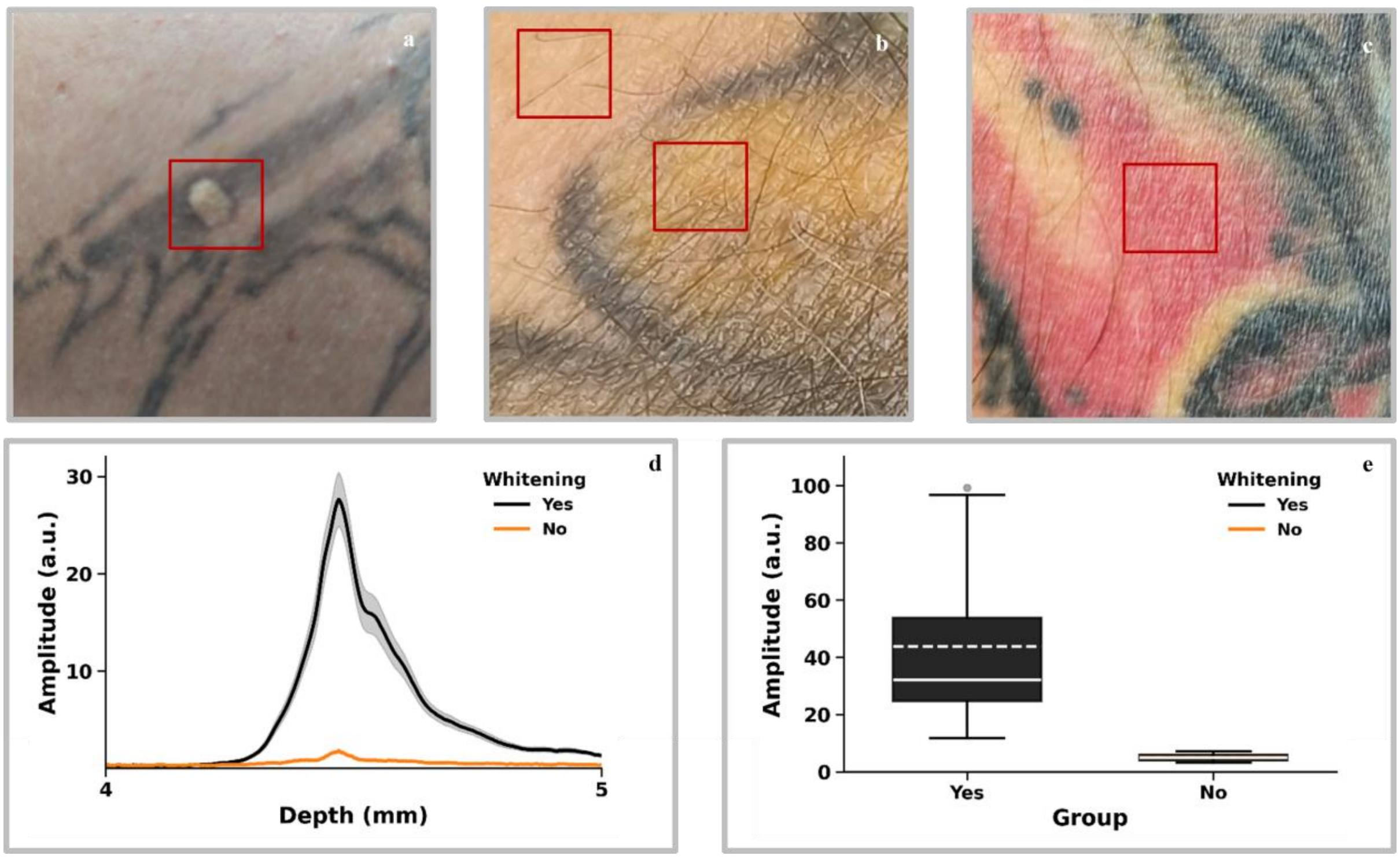

**Figure 3**: Representative photographic pictures of different zones after treatment: (**a**) black ink (whitening), (**b**) yellow ink and untreated skin (no whitening), and (**c**) red ink (no whitening). Red boxes mark the analysed regions. (**d**) Mean aligned Hilbert envelope amplitude profiles versus depth derived from all baseline optoacoustic measurements within two whitening categories. Black ink zones, associated with whitening, showed higher amplitudes, whereas yellow/red ink and skin zones, associated with no visible whitening, showed lower (almost 0 a.u.) amplitudes. Shaded areas indicate the standard error of the mean (SEM). (**e**) Boxplot of mean P2P values for both whitening categories, showing higher optoacoustic amplitudes in zones with whitening. The two boxplots are clearly separated and show no overlap. The solid line inside each box indicates the median, while the dashed line represents the mean.

**Clinical study 2.**

Figure 4 shows the longitudinal evolution of the optoacoustic response after laser treatment of different zones. Photographic pictures from one representative zone (zone 1) are presented in Figure 4a-e, together with the corresponding raw optoacoustic signals and Hilbert envelopes in Figure 4f-j. Due to space limitations, only the results from one zone, acquired before treatment and at 3, 9, 15, and 21 min after treatment, are shown. The photographic pictures showed a clear immediate whitening response at 3 min after treatment within the irradiated area, followed by a progressive reduction in the visible whitening effect at later time points, with only minimal tattoo pigment removal being visually observed within the treated region.

The temporal evolution of the whitening was consistent with the optoacoustic measurements. Before treatment, the zone exhibited a well-defined optoacoustic response. At 3 min after laser exposure, both the raw signal and the Hilbert envelope were markedly attenuated, indicating a strong early suppression of the detectable response due to the whitening-related light scattering, which prevents the excitation light to reach the tattoo. At 9-12 min, the signal reappeared with substantially higher amplitude and this recovery was maintained over time, with a well-defined peak.

Figure 4k highlights the temporal behaviour of the normalized P2P amplitude across zones after laser treatment. For visualization, the curves were smoothed using PCHIP interpolation. In general, the measurements showed a pronounced early drop shortly after treatment, followed by a variable degree of recovery over time. The degree of optoacoustic signal recovery was inversely related to the degree of visible tattoo pigment removal within the treated region. The zone showing the lowest tattoo pigment removal exhibited the strongest recovery of the optoacoustic response (zone 1). As tattoo pigment removal progressively increased from zones 2 to 3, P2P amplitudes became lower over time. The zone showing the highest tattoo pigment removal exhibited the lowest P2P amplitudes (zone 4) and a sustained suppression of the optoacoustic signal throughout the observation period. This interpretation was consistent with the photographic pictures, in which the treated regions showed different degrees of tattoo pigment removal at later time points, increasing progressively from zone 1 to zone 4. This was particularly evident in zone 4, where tattoo pigment was barely observed within the treated region at 30 min, consistent with extensive tattoo pigment removal in the analysed area (see “Supplementary Note 1”). This interpretation was further supported by clinical follow-up 3 months after treatment, which confirmed the absence of most tattoo pigment within the treated region.

The analysis of the mean P2P values and the degree of ink removal showed clear results. The tattoo pigment removal percentages calculated for the four treated zones were 19.36 %, 59.13 %, 71.66 %, and 83.30 %, respectively (see Supplementary Note 2). The corresponding mean P2P values at 30 min were 29.45 ± 7.99 a.u., 11.63 ± 4.56 a.u., 6.99 ± 1.30 a.u., and 4.53 ± 0.82 a.u., respectively. Thus, zones with higher tattoo pigment removal percentages showed lower mean optoacoustic amplitudes over time. This relationship is shown in Figure 4l, where linear regression revealed a negative association between tattoo pigment removal percentage and mean P2P amplitude ($R^2 = 0.82$, $p < 1 \times 10^{-6}$).

In summary, these results indicate that the optoacoustic signal decay right after laser treatment since the increase scattering due to whitening does not allow the light to reach the tattoo. However, as time passes (and as long as the ink remains in place) the optoacoustic response recover due to the progressive fading of the whitening effect which also the light to reach the tattoo again.

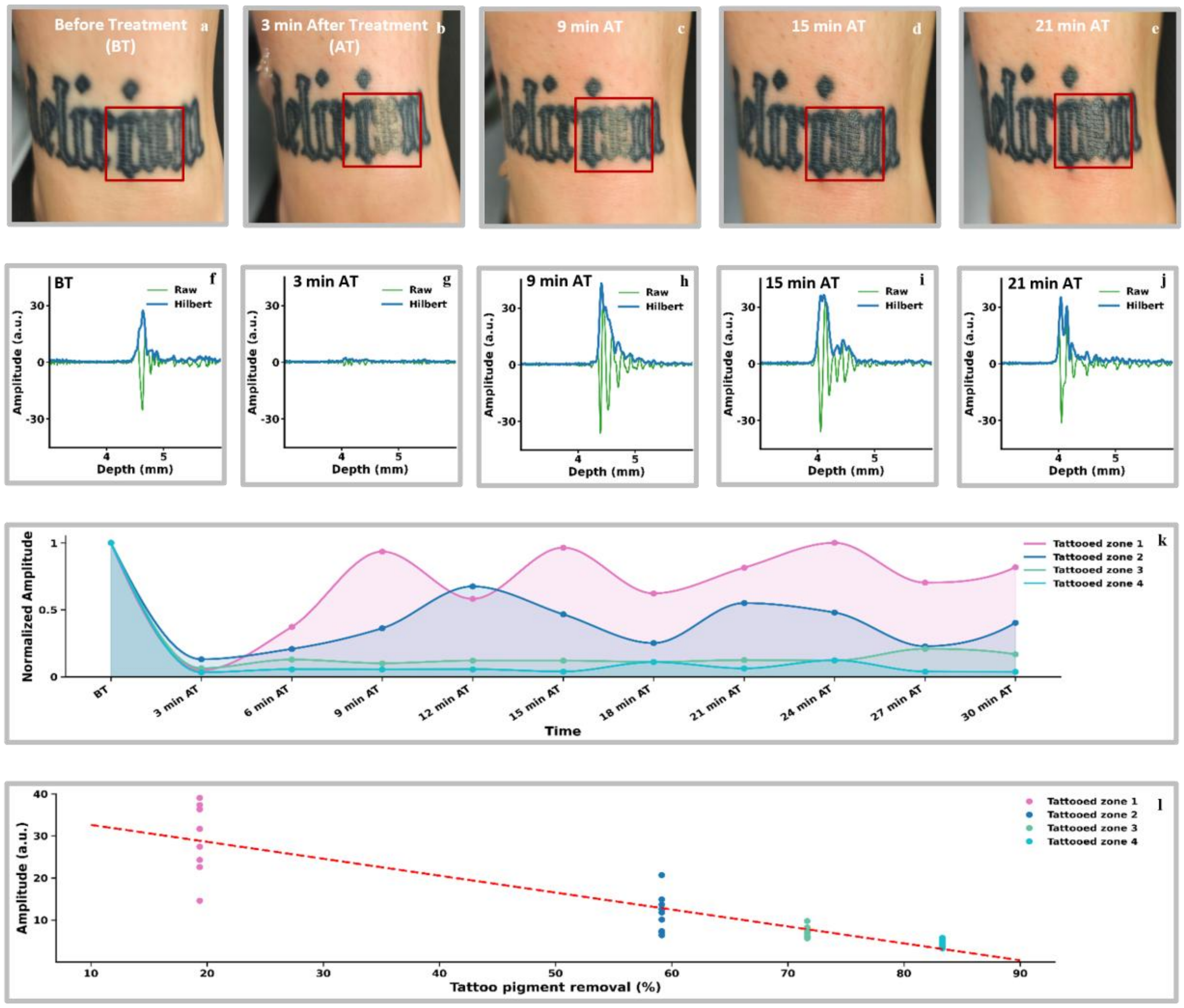

**Figure 4**: Longitudinal clinical pilot study and optoacoustic monitoring following laser treatment. (**a–e**) Representative photographic pictures of the same zone acquired before treatment (BT) and at 3, 9, 15, and 21 min after treatment (AT). Red squares mark the region treated and analysed with the SOL system. (**f–j**) Corresponding optoacoustic waveforms, showing the raw signal (green) and its Hilbert envelope (blue) at each time point. (**k**) Time course of the normalized peak-to-peak (P2P) amplitude for all zones before treatment (BT) to 30 min after treatment (30 min AT). Curves were smoothed using PCHIP interpolation for visualization. Colours indicate the different zones. (**l**) Relationship between tattoo pigment removal percentage and mean P2P amplitude. Each point corresponds to one mean P2P measurement acquired between 6 and 30 min after treatment and paired with the tattoo pigment removal percentage calculated for the corresponding treated zone. The dashed red line represents the fitted linear regression model.

## Discussion

We have studied the relationship between OI and SP both theoretically and through in-human experiments on tattoo removal. The results show a strong correlation between the intensity of the optoacoustic images and the immediate clinical endpoint. Analysis of the results indicates that OI systems such as SOL may have a substantial positive impact on SP procedures in general, and on tattoo removal procedures in particular.

The direct relationship between the strength of the optoacoustic signals before treatment and the level of whitening indicates that SOL has the potential to be used as a dosimetry tool for the rapid selection of the optimal wavelength. For multi-coloured tattoos displaying many different zones whose absorption spectra may be unknown, the standard trial-and-error method for identifying the optimal wavelength might be too lengthy or even impractical, especially for SP systems with 4 or 5 accessible wavelengths. A multispectral SOL system could indentify the optimal wavelength for each zone in less than a second. The theoretical analysis also supports the validity of this application.

The direct relationship between the strength of the optoacoustic signals before treatment and the degree of whitening also indicates that, after a larger sample-size study, the SOL system could be calibrated to provide an initial estimate of the optimal energy per pulse, which could then be adjusted through the standard trial-and-error process. In this way, energy selection time is reduced. Less experienced SP practitioners would greatly benefit from the SP information, shortening the learning curve. This would be a data-driven approach, suitable for example for AI algorithms, since the theoretical analysis indicates that, while there is a direct relation between optoacoustic images and energy selection in SP, a distinct analytical expression cannot be derived.

The relationship between the optoacoustic signals after treatment and the level of whitening also indicates strong clinical potential for reducing the R20 treatment time. In the current recommended practise, if a dermatologist wants to perform a second laser pass within the same session, he must wait 20 minutes to ensure that the scattering caused by whitening has returned to baseline levels. However, we observed that optoacoustic signals can return to baseline levels as soon as 9-12 min, indicating light absorption by the tattoo and the possible early application of the next SP treatment pass. The fact that, in clinical study 1 we were able to correlate the optoacoustic signal with the clinical endpoint supports our hypothesis, which is further supported by the theoretical analysis.

Many other applications of OI systems and particularly SOL in tattoo removal can be envisioned and should be investigated in the near future. For example, the morphological imaging capabilities of SOL could be used to determine the depth of the tattoo and adjust the spot size accordingly, which, ideally, should be selected as a function of tattoo depth[2]. Similarly, SOL images could be used to obtain a precise estimate of the treatment effect on the tattoo and to better decide whether another treatment session is worthwhile.

The dosimetric capabilities of OI can also lead to other applications. For example, it could be used to determine the amount of light that will be absorbed by the melanin layer and by tattoo during treatment. This information may be crucial for selecting the optimal wavelength and making better-informed treatment decisions. Standard observation of whitening might not be sufficient to determine whether the whitening is dermal or epidermal[4].

Lasers and intense pulsed light systems are used in dermatology settings to treat several diseases including vitiligo, rosacea and port-wine stains. In such applications, the laser pulse duration tends to be significantly larger than in OI to ensure appropriate heating of the targets. In such cases, the OI can provide a map of the initial distribution of the absorbed heat.

Further work should investigate the possibility of using the SOL system as an auxiliary tool for SP age-spot removal. Age-spot elimination is often performed with the same system used for tattoo removal. Therefore, some of the advantages of the SOL system for tattoo removal may apply for aging mark removal.

This work paves the way for using optoacoustic imaging in general, and the SOL system in particular, to substantially improve the safety and precision of tattoo removal. The principles discussed here can be extended, perhaps with small modifications, to the broad range of cosmetic and medical treatments based on selective photothermolysis.

## Data and Code Availability

Code and data generated during the current study are available from the corresponding author upon reasonable request.

## Author Contribution Statement

Conceptualization, J.A.; methodology, J.A.; software, S.C., R.R. and J.A.; validation, S.C.; formal analysis, S.C.; investigation, S.C., A.J. and R.R.; resources, J.A.; data curation, S.C.; writing—original draft preparation, J.A. and S.C.; writing—review and editing, J.A. and S.C.; visualization, S.C.; supervision, J.A.; project administration, J.A.; funding acquisition, J.A. All authors have read and agreed to the published version of the manuscript.

## Acknowledgements

This research was funded by the Madrid Talento research grant 2020-T1/TIC-20661 (PRECISION).

## Competing Interests

Juan Aguirre is a founder and equity owner of the company Hypersensors-Medical which, however, did not fund the study. Sergio Contador and Juan Aguirre are co-authors of the patent EP25382275, which protects several embodiments of the SOL system.

## Ethics Approval and Informed Consent

This study was conducted in accordance with the Declaration of Helsinki and was approved by Ethical Committee of the Autonomous University of Madrid under approval number CEI-148- 3460 . All participants provided written informed consent before inclusion in the study, including consent for optoacoustic measurements, laser treatment and photographic documentation.